

\input lanlmac

\def\d{\partial}

\def\o{\over}
\def\hf{{1 \over 2}}
\def\bra#1{\left\langle #1 \right|}
\def\ket#1{\left| #1 \right\rangle}
\def\bvac{\left\langle0\right|}
\def\V{\left\langle V_3 \right|}
\def\Vt{\left\langle \tilde{V}_3 \right|}
\def\sitarel#1#2{\mathrel{\mathop{\kern0pt #1}\limits_{#2}}}
\def\a{\alpha}
\def\A{\Psi}

\def\np#1#2#3{{ Nucl. Phys.} {\bf B#1}, #2 (19#3)}

\def\pln#1#2#3{{Phys. Lett. } {\bf B#1}, #2 (19#3)}
\def\plo#1#2#3{{ Phys. Lett.} {\bf #1B}, #2 (19#3)}
\def\pr#1#2#3{{ Phys. Rev.} {\bf D#1}, #2 (19#3)}
\def\prl#1#2#3{{ Phys. Rev. Lett.} {\bf #1}, #2 (19#3) }

\def\ann#1#2#3{{ Ann. Phys.} {\bf #1}, #2 (19#3)}
\def\ptp#1#2#3{{ Prog. Theor. Phys.} {\bf #1}, #2 (19#3)}

\def\mpl#1#2#3{{ Mod. Phys. Lett.} {\bf A#1}, #2 (19#3)}
\def\jhep#1#2#3{{JHEP} {\bf #1}, #2 (19#3)}

\def\hpt#1{{\tt hep-th/#1}}

\lref\CDS{
A.~Connes, M.R.~Douglas and A.~Schwarz, 
``Noncommutative Geometry and Matrix Theory: Compactification on Tori,''
\jhep{02}{003}{98}, \hpt{9711162}.
}
\lref\DH{
M.R.~Douglas and C.~Hull,
``D-Branes and the Noncommutative Torus,''
\jhep{02}{008}{98}, \hpt{9711165}\semi
Y.-K.E. Cheung and M. Krogh,
``Noncommutative Geometry from 0-branes in a Background B Field,''
\np{528}{185}{98}, \hpt{9803031}\semi
T. Kawano and K. Okuyama,
``Matrix Theory on Noncommutative Torus,''
\pln{433}{29}{98}, \hpt{9803044}.
}
\lref\Schom{
V. Schomerus,
``D-branes and Deformation Quantization,''
\jhep{9906}{030}{99}, \hpt{9903205}.
}
\lref\SW{
N.~Seiberg and E.~Witten, 
``String Theory and Noncommutative Geometry,''
\hpt{9908142}.
}
\lref\AsaKis{
T.~Asakawa and I.~Kishimoto,
``Comments on Gauge Equivalence in Noncommutative Geometry,''
\jhep{9911}{024}{99}, \hpt{9909139}.
}
\lref\SWrel{
L.~Cornalba, 
``D-brane Physics and Noncommutative Yang-Mills Theory,''
\hpt{9909081}\semi
N.~Ishibashi, 
``A Relation between Commutative and Noncommutative Descriptions of D-branes,''
\hpt{9909176}\semi
K.~Okuyama, 
``A Path Integral Representation of the Map between
Commutative and Noncommutative Gauge Fields,''
\hpt{9910138}.
}
\lref\WSFT{
E.~Witten, 
``Noncommutative Geometry and String Field Theory,''
\np{268}{253}{86}.
}
\lref\HIKKO{
H.~Hata, K.~Itoh, T.~Kugo, H.~Kunitomo and K.~Ogawa,
``Covariant String Field Theory,''
\pr{34}{2360}{86}.
}
\lref\Sen{
A.~Sen,
``Open String Field Theory in Nontrivial Background Field (I): 
Gauge Invariant Action,''
\np{334}{350}{90}.
}
\lref\MS{
T.R. Morris and B. Spence,
``Background Independent String Field Theory,''
\np{316}{113}{89}.
}
\lref\HLRS{
G.T.~Horowitz, J.~Lykken, R.~Rohm and A.~Strominger,
``Purely Cubic Action for String Field Theory,''
\prl{57}{283}{86}.
}
\lref\HIKKOpre{
H.~Hata, K.~Itoh, T.~Kugo, H.~Kunitomo and K.~Ogawa,
``Pregeometrical String Field Theory: Creation of Space-Time and Motion,''
\plo{175}{138}{86}.
}
\lref\ChuHo{
C.-S.~Chu and P.-M.~Ho, 
``Noncommutative Open String and $D$-Brane,''
\np{550}{151}{99}, \hpt{9812219};
``Constrained Quantization of Open String in Background $B$ Field and 
Noncommutative $D$-Brane,'' \hpt{9906192}.
}
\lref\SJabbari{
F.~Ardalan, H.~Arfaei, and M.~M.~Sheikh-Jabbari, 
``Dirac Quantization of Open Strings and Noncommutativity in Branes,'' 
\hpt{9906161}\semi 
M.~M.~Sheikh-Jabbari and A.~Shirzad, 
``Boundary Conditions as Dirac Constraints,''
\hpt{9907055}.
}
\lref\HS{
G.T.~Horowitz and A.~Strominger,
``Translations as Inner Derivations and Associativity Anomaly in 
Open String Field Theory,'' \pln{185}{45}{87}.
}
\lref\LPP{
A.~Leclair, M.~Peskin and C. R.~Preitschopf,
``String Field Theory on the Conformal Plane (I),''
\np{317}{411}{89}.
}
\lref\Samuel{
S.~Samuel, 
``The Physical and Ghost Vertices in Witten's String Field Theory,''
\plo{181}{255}{86}.
}
\lref\GJ{
D.J.~Gross and A.~Jevicki,
``Operator Formulation of Interacting String Field Theory (I),''
\np{283}{1}{87};
``Operator Formulation of Interacting String Field Theory (II),''
\np{287}{225}{87};
``Operator Formulation of Interacting String Field Theory (III), 
NSR Superstring,''
\np{293}{29}{87}.
}
\lref\CST{
E.~Cremmer, A.~Schwimmer and C.~Thorn,
``The Vertex Function in Witten's Formulation of String Field Theory,''
\plo{179}{57}{86}.
}
\lref\Yoneya{
T.~Yoneya,
``String Coupling Constant and Dilaton Vacuum Expectation Value in String 
Field Theory,''
\pln{197}{76}{87}.
}
\lref\FS{
M.~Fisk and M.~Srendnicki,
``Magnetic String Fields,''
\np{313}{308}{89}.
}
\lref\WittenSSFT{
E.~Witten,
``Interacting Field Theory of Open Superstrings,''
\np{276}{291}{86}.
}
\lref\Suehiro{
K.~Suehiro,
``Operator Expression of Witten's Superstring Vertex,''
\np{296}{333}{88}.
}
\lref\Wendt{
C.~Wendt,
``Scattering Amplitudes and Contact Interactions in
Witten's Superstring Field Theory,''
\np{314}{209}{89}.
}
\lref\KugoZwie{
T.~Kugo and B.~Zwiebach,
``Target Space Duality as a Symmetry of String Field Theory,''
\ptp{87}{801}{92}.
}
\lref\Berkovits{
N.~Berkovits and C.T.~Echevarria,
``Four-Point Amplitude from Open Superstring Field Theory,''
\hpt{99012120}\semi
N.~Berkovits,
``A New Approach to Superstring Field Theory,''
\hpt{99012121}.
}
\lref\Sugino{
F.~Sugino,
``Witten's Open String Field Theory in Constant $B$-field Background,''
\hpt{9912254}.
}
\lref\Wopenclosed{
B.~Zwiebach, 
``Oriented Open-Closed String Field Theory Revisited,''
\ann{267}{193}{98}, \hpt{9705241}; 
``Quantum Open String Theory with Manifest Closed String Factrization,''
\pln{256}{22}{91};
``Interpolating String Field Theories,'' 
\mpl{7}{1079}{92}, \hpt{9202015}.
}
\lref\AKT{
T.~Asakawa, T.~Kugo, and  T.~Takahashi,
``One-Loop Tachyon Amplitude in Unoriented Open-Closed String Field Theory,''
\ptp{102}{427}{99}, \hpt{9905043};
``BRS Invariance of Unoriented Open-Closed String Field Theory,''
\ptp{100}{831}{99}, \hpt{9807066}\semi
T.~Kugo and  T.~Takahashi,
``Unoriented Open-Closed String Field Theory,''
\ptp{99}{649}{98}, \hpt{9711100}.
}
\lref\KT{
T.~Kawano and T.~Takahashi,
in preparation.
}
\lref\Polchn{
J.~Polchinski, 
``Dirichlet Branes and Ramond-Ramond Charges,''
\prl{75}{4724}{95}, \hpt{9510017}.
}
\lref\DBI{
R.G.~Leigh,
``Dirac-Born-Infeld Action from Dirichlet Sigma Model,''
\mpl{4}{2726}{89}.
}


\Title{                                \vbox{\hbox{UT-870}
                                             \hbox{\tt hep-th/9912274}} }
{\vbox{\centerline{
                  Open String Field Theory on Noncommutative Space
}}}

\vskip .2in

\centerline{
                       Teruhiko Kawano and Tomohiko Takahashi
}

\vskip .2in 


\centerline{\sl
                      Department of Physics, University of Tokyo
}
\centerline{\sl
                            Hongo, Tokyo 113-0033, Japan
}
\centerline{\tt
                         kawano@hep-th.phys.s.u-tokyo.ac.jp
}
\vskip -1mm
\centerline{\tt
                          tomo@hep-th.phys.s.u-tokyo.ac.jp
}

\vskip 3cm

We study Witten's open string field theory 
in the presence of a constant $B$ field. 
We construct the string field theory in the operator formalism and find that, 
compared to the ordinary theory with no $B$ field, 
the vertices in the resulting theory has an additional factor.
The factor makes the zero modes of strings noncommutative. 
This is in agreement with the results in the first-quantized formulation.
We also discuss background independence of the purely cubic action derived 
from the above string field theory and then find a redefinition of string 
fields to remove the additional factor from the vertex.
Furthermore, we briefly discuss the supersymmetric extension of our string 
field theory.

\bigskip
\Date{December, 1999}


\newsec{Introduction}

Since the appearance of the seminal paper \CDS, 
noncommutative geometry has received much attention in Matrix theory and 
string theory \refs{\DH,\SW}. See \SW\ for further references.
In string theory, we have the familiar antisymmetric tensor field $B_{ij}$ 
which directly couples to fundamental strings. If we turn on the background 
$B$ field, spacetime becomes noncommutative on $D$-branes with 
the nonvanishing $B$ field. $D$-branes can be described by open strings whose 
ends are on the $D$-branes \Polchn. By the quantization of the open strings, 
we have gauge field on the $D$-branes, and the low-energy effective theory 
of the gauge field are described by the Dirac-Born-Infeld (DBI) action \DBI. 
Therefore, turning on the $B$ field, we can find the DBI action 
on the noncommutative space. 

Recently, Seiberg and Witten have shown that the noncommutative DBI action is 
equivalent to the ordinary one \SW. To prove the equivalence, they have given 
a relation between the gauge fields in the noncommutative DBI action and 
the ordinary one \SW. Some closely related topics have been discussed in 
\refs{\AsaKis,\SWrel}. However, at present it seems unclear how we can embed 
the relation into a whole tower of the excitation modes of strings. 
To uncover such a relation, string field theories seems a natural framework, 
where we can deal with string fields which include all the excitations as well 
as the gauge field.

In the paper \WSFT, Witten has constructed, on a commutative flat Minkowski 
spacetime, a covariant open string field theory based on noncommutative 
geometry. This noncommutativity comes from the nature of the way that 
open strings join together to become a new string. 
Therefore, we may expect that Witten's string field theory in the background 
$B$ field has additional noncommutativity. In this paper, we will derive 
Witten's open string field theory in the above-mentioned background 
in the operator formalism \refs{\CST,\Samuel,\GJ}. 
By solving the overlap conditions, 
we will show that the string field theory has an additional factor in 
its vertex. This factor accounts for the noncommutativity of spacetime and 
is in agreement with the result of \refs{\Schom,\SW} in the first-quantized 
formulation. In \FS, this factor has also been found in Witten's open string 
field theory with a constant background magnetic field $F_{ij}$. By the gauge 
invariance $B_{ij}\rightarrow B_{ij}+\d_i\Lambda_j-\d_j\Lambda_i$, 
$A_i \rightarrow A_i+\Lambda_i$, we can see that this background is the same as 
ours. However, the physical significance of this factor has not been fully
realized. Also, open string field theories in general backgrounds have been 
discussed in terms of the conformal field theory \Sen. 
The string field theory in this paper could be studied in the same way. 

Pregeometrical string field theories have been proposed 
to give a background independent formulation of 
string theory \refs{\HLRS,\HIKKOpre}. 
In particular, the pregeometrical theory given in \HLRS\ is 
Witten's open string field theory on a flat Minkowski spacetime 
without the kinetic term and so it is sometimes referred to 
as purely cubic action. 
Therefore, if we drop the kinetic term from our string theory 
in the background $B$ field, it is tempting to ask whether the resulting theory 
can be background independent. Since the additional noncommutative factor 
explicitly depends on the background $B$ field, we may at first think that 
it cannot be background independent. If we dealt with a particle field 
theory, this would be true. However, as we will show in this paper, 
we can remove 
the noncommutative factor from the three-string vertex by a redefinition of 
string fields. In addition, we will explicitly demonstrate that 
the three-string vertex is independent of the background metric which we use 
to express the vertex in terms of the oscillators of strings. To this end, we
will apply the method given in \KugoZwie\ to open string field theory.

This paper is organized as follows: 
In section 2, we construct Witten's open string field theory 
in the background $B$ field by using the operator formalism and solving 
the overlap conditions for the vertices. 
In section 3, we explicitly show background independence of 
our pregeometrical theory in great detail.
Section 4 is devoted to discussion. In appendix A, we briefly summarize 
the operator formalism of the first-quantized string theory 
\refs{\ChuHo,\SJabbari}. 
In appendix B, we give a derivation of Yoneya's identities \refs{\GJ,\Yoneya} 
of the Neumann coefficients for Witten's string field theory, which we need 
in section 3.

When we had almost finished writing this paper, we found a paper \Sugino\ 
given by Sugino, 
which has considerable overlap with ours. 
The main difference between that paper and ours 
is the following two points. First, he argues that the dependence on the 
$B$ field can be eliminated from our string field theory by a redefinition of 
string fields. This suggests that we can `gauge away' the background 
field. We will discuss this point in further detail in section 4. 
Second, we explicitly show background independence of our pregeometrical 
theory.  In section 4, we will also mention our main results about 
an open-closed string field theory with the light-cone type interaction \AKT\ 
in the background we are considering. Furthermore, 
we will discuss the supersymmetric extension \WittenSSFT\ 
of our string field theory.


\newsec{String Field Theory in Background $B$-Field}

We study a bosonic open string field theory proposed by Witten \WSFT\ 
with a constant metric $g_{ij}$ and a constant antisymmetric field $B_{ij}$.
The open string field theory in the presence of background fields
has been discussed in \refs{\Sen,\MS}. We show that we can construct 
the field theory in our background explicitly 
by using the operator formalism \refs{\CST,\Samuel,\GJ}. 
To this end, it is appropriate to begin with a review of the operator 
formalism of the first-quantized string theory with the $B$ field 
\refs{\ChuHo,\SJabbari}. 
In appendix A, we give a simple derivation of the result given by 
\refs{\ChuHo,\SJabbari} to make this paper self-contained.

In the first-quantized string theory, the worldsheet action is given by
\eqn\WSaction{
S=-{1\over4\pi\alpha'}\int d\sigma\,d\tau
\left(g_{ij}\eta^{ab}\partial_a X^i \partial_b X^j
-2\pi\alpha'B_{ij}\epsilon^{ab}
\partial_a X^i \partial_b X^j \right),
}
From this action, if the Dirichlet boundary 
condition is not chosen for all the directions of the string coordinates, 
the boundary condition 
can be seen to be $g_{ij}{X^j}'+ (2\pi\alpha') B_{ij}\dot{X}^j =0$ 
at $\sigma=0,\,\pi$, where we denote the differentiation with respect to 
$\tau$ and $\sigma$ by the dot $\ \dot{}$ and the prime ${}'$, respectively.
For simplicity, in this paper, we impose this boundary condition 
on all the string coordinates $X^i(\tau,\sigma)$.
The conjugate momenta of the string coordinates $X^i(\sigma)$ turn out to be
$P_i(\sigma) = {1\over2\pi\alpha'}g_{ij}\dot{X}^j(\sigma) + B_{ij}
{X^j}'(\sigma)$.

The authors of \refs{\ChuHo,\SJabbari} have shown that 
we can quantize our system by the Dirac 
quantization procedure if we treat the boundary condition 
as a constraint. See also appendix A for further details.
The resulting commutation relations can be seen to be
\eqn\CCR{\eqalign{
&\left[X^i(\sigma),\ P_j(\sigma')\right]=i\,
\delta^i_j\,\delta(\sigma-\sigma'), 
\cr
&\left[P_i(\sigma),\ P_j(\sigma')\right]=0, 
\cr
&\left[X^i(\sigma),\ X^j(\sigma')\right]=\left\{
\matrix{
&\ i\theta^{ij}, &(\sigma=\sigma'=0)\cr
&-i\theta^{ij}, &(\sigma=\sigma'=\pi)\cr
&0, &({\rm otherwise}),
}
\right.
\cr
}}
where we use the same definitions of the open string metric $G_{ij}$ and 
the theta parameter $\theta^{ij}$ as those in \SW:
\eqn\openmetric{\eqalign{
G^{ij}&=\left({1 \over g+2\pi\alpha'B}g{1 \over g-2\pi\alpha'B}
\right)^{ij}, 
\cr
\theta^{ij}&=-(2\pi\alpha)^2
\left({1 \over g+2\pi\alpha'B}B{1 \over g-2\pi\alpha'B}
\right)^{ij}.
\cr}
}

As we can see from appendix A, the mode expansion of the string coordinates
$X^i(\sigma)$ turns out to be 
\eqn\Xmode{
X^i(\sigma)=\tilde{X}^i(\sigma)+{(\theta G)^i}_j Q^j(\sigma), 
}
where $\tilde{X}^i(\sigma)$ and $Q^i(\sigma)$ are defined with 
$l_s=\sqrt{2\a'}$ as follows:
\eqn\XQmode{\eqalign{
&\tilde{X}^i(\sigma)=\tilde{x}^i 
+l_s\sum_{n\neq0}{i \over n}\a_n^i\cos(n\sigma), 
\cr
&Q^i(\sigma)={1\o\pi}G^{ij}p_j\left(\sigma-{\pi\o2}\right)
+{1\o\pi l_s}\sum_{n\neq0}{1 \o n}\a_n^i\sin(n\sigma).
\cr}
}
We can see here that the variables $\tilde{X}^i(\sigma)$ satisfy the Neumann 
boundary condition. Similarly, the expansion of the momenta is
\eqn\Pmode{
P_i(\sigma)={1\o\pi l_s}\sum_nG_{ij}\a_n^j\cos(n\sigma).
}
Note that $Q^i(\sigma)$ and $P_i(\sigma)$ have
the following relation: 
\eqn\qp{
Q^i(\sigma)=\int_{\pi\o2}^\sigma d\sigma' G^{ij}P_j(\sigma')
+\hf G^{ij}\left(p_{L\,j}-p_{R\,j}\right),
}
where we introduce the momentum operators integrated
over half a string \HS,
\eqn\halfP{
p_{L\,i}=\int_0^{\pi\o2} d\sigma\ P_i(\sigma),
\qquad
p_{R\,i}=\int_{\pi\o2}^\pi d\sigma\ P_i(\sigma).
}
The commutation relations of these mode variables $\tilde{x}^i$, $p_i$, 
$\a^i_n$ can be verified \refs{\ChuHo,\SJabbari} to be
\eqn\commutator{
[\tilde{x}^i,\,p_j]=i{\delta^i}_j,
\qquad
\left[\a_m^i,\,\a_n^j\right]=m\delta_{m+n}G^{ij},
}
and the others vanish, as is seen from appendix A.

The BRS charge in string field theories is necessary 
to construct their kinetic terms. 
In order to obtain the BRS charge, we need to know 
the energy-momentum tensor in the worldsheet theory \WSaction. 
While the contribution from the reparametrization ghosts to the 
energy-momentum tensor is the same 
as usual, the energy-momentum tensor from the matter sector, 
{\it i.e.} the string coordinates, can be found to be
\eqn\EMtensor{\eqalign{
&T(z) = -{1\o\alpha'}\,g_{ij}\partial X^i\partial X^j(z)
=-{1\o\alpha'}\,G_{ij}\partial \tilde{X}^i\partial \tilde{X}^j(z),
\cr
&\tilde{T}(\bar{z}) = -{1\o\alpha'}\,g_{ij}\bar{\partial}
 X^i\bar{\partial} X^j(\bar{z})=
 -{1\o\alpha'}\,G_{ij}\bar{\partial}
 \tilde{X}^i\bar{\partial} \tilde{X}^j(\bar{z}),
\cr}
}
with $z=\exp(\tau+i\sigma)$. 
Note from the boundary condition that $T(z)=\tilde{T}(\bar{z})$ for $z=\bar{z}$.
Therefore, we can make use of the doubling technique for open strings to extend 
the worldsheet of the upper half-plane to a whole complex plane when 
we define the BRS charge by using the energy momentum tensor 
in the usual way.

In the rest of this section, we will construct a string field theory 
with the mid-point interaction in our background. To this end, the reflector 
and the three-string vertex will be constructed by using 
the overlap conditions, as usual. From the paper \refs{\Schom,\SW}, 
we expect that 
the noncommutativity of spacetime would also appear in our string field theory, 
in addition to the usual noncommutativity of an open string field theory. 
This is actually the case, as we will see below. 
Although the mode expansion of the string coordinates is different from 
that with the Neumann boundary condition due to 
the presence of the background $B$-field, the resulting vertices will be shown 
to be the same as usual vertices except for one factor. It is this factor that 
accounts for the noncommutativity of spacetime. By noncommutative  
spacetime, we mean a spacetime such that, given two arbitrary functions $f(x)$ 
and $g(x)$ on the space, the product of these functions is given by 
the Moyal product
\eqn\starprd{
f*g(x)=
f(x)\exp\left[{i\o2}\theta^{ij}\overleftarrow\d_i\overrightarrow\d_j\right]g(x).
}
If we identify the `zero mode' $\tilde{x}$ of string coordinates with 
coordinates of our spacetime, the above-mentioned factor turns out to be the 
exponential factor in \starprd, as we will show below.

Besides the BRS charge, in order to obtain a kinetic term in string field 
theory, we need the reflector $\left\langle R\right|$, which is used to give 
the inner product of string fields.
The reflector $\left\langle R\right|$ is defined up to an overall 
normalization by the overlap conditions
\eqn\connection{\eqalign{
\left\langle R\right|\left({X^i}^{(1)}(\sigma)-{X^i}^{(2)}
(\pi-\sigma)\right)=0, 
\cr
\left\langle R\right|\left(P_j^{(1)}(\sigma)+P_j^{(2)}
(\pi-\sigma)\right)=0.
\cr}
}
Since the ghost part of the reflector remains unchanged even in our case, 
we will focus on only the matter part of it. 
This will also be the case later for the three-string vertex.
In a case where $\theta=0$, the matter part of the reflector is thus given by
\eqn\reflector{
\left\langle R^x \right| = (2\pi)^{26}\delta^{26}(p_1+p_2)
{}_{21}\left\langle0\right|
\exp\left(-\sum_{n\geq 1}
{(-)^n \o n}G_{ij}{\a^i_n}^{(1)}{\a^j_n}^{(2)}\right),
}
where ${}_{21}\left\langle0\right|$ denotes
${}_2\left\langle0\right|{}_1\left\langle0\right|$.
We can see that this reflector still satisfies the connection conditions 
\reflector\ even in our case of $\theta\neq 0$.
Therefore, using the BRS charge $Q_{\rm B}$ and the reflector 
$\left\langle R\right|$, 
we can write the kinetic term of our string field theory. 
Obviously, this kinetic term does not have any dependence of the 
theta parameter $\theta^{ij}$.

Now, let us move on to the three-string vertex.
The three-string vertex can also be specified up to an overall normalization by 
the connection equations
\eqn\VXP{\eqalign{
&\V \left({X^i}^{(r)}(\sigma)-{X^i}^{(r+1)}(\pi-\sigma)
\right)=0,\ \ \ ({\pi\o2}< \sigma \leq \pi), 
\cr
&\V \left({P_i}^{(r)}(\sigma)+{P_i}^{(r+1)}(\pi-\sigma)
\right)=0,\ \ \ ({\pi\o2}< \sigma \leq \pi),
\cr}
}
where $r=1,2,3$ denotes the $r$-th string and $r+3$ equals $r$.
Before proceeding to solve these equations, let us consider the three-string 
vertex with $\theta=0$. 
\eqn\VtX{\eqalign{
&\Vt \left(\tilde{X}^{i\,(r)}(\sigma)-\tilde{X}^{i\,(r+1)}(\pi-\sigma)
\right)=0,\ \ \ ({\pi\o2}< \sigma \leq \pi) 
\cr
&\Vt \left({P_i}^{(r)}(\sigma)+{P_i}^{(r+1)}(\pi-\sigma)
\right)=0.\ \ \ ({\pi\o2}< \sigma \leq \pi)
\cr}
}
The variables $\tilde{X}^i(\sigma)$ and $P_i(\sigma)$ are expressed
by the mode expansions of \Xmode\ and \Pmode, and these correspond
to strings with the Neumann boundary condition as mentioned above.
Therefore, we find that this vertex $\Vt$ agrees with the usual
three-string vertex given by 
\eqn\VVV{\eqalign{
&\left\langle \tilde{V}_3^x\right|
=(2\pi)^{26}\delta^{26}(\sum_{r=1}^3 p^{(r)})
{}_{321}\left\langle0\right| e^{E_{123}}, 
\cr
&E_{123}=
\sum_{\sitarel{m,n\geq 0}{r,s=1,2,3}}
\hf\bar{N}_{mn}^{rs}
G_{ij}{\a_m^i}^{(r)}{\a_n^j}^{(s)},
\cr}
}
where the Neumann coefficients share the same forms as those in the 
Minkowski spacetime \refs{\CST,\Samuel,\GJ} and ${}_{321}\left\langle0\right|$ 
denotes ${}_{3}\left\langle0\right|{}_{2}\left\langle0\right|{}_{1}
\left\langle0\right|$.

Using \qp, \Xmode, and \VtX, 
we can evaluate how the string coordinates ${X^i}^{(r)}(\sigma)$ connect
with each other on the vertex $\Vt$
\eqn\newconct{
\Vt \left({X^i}^{(r)}(\sigma)-{X^i}^{(r+1)}(\pi-\sigma)\right)
= -\hf\Vt \theta^{ij} p_j^{(r+2)},\ \ \ ({\pi\o2}< \sigma \leq \pi)
}
where, due to the momentum conservation on the worldsheet \HS,  
\eqn\PLPR{
p_{L\,i}^{(r)}+p_{R\,i}^{(r)}=p_i^{(r)},\ \ \
p_{L\,i}^{(r+1)}+p_{R\,i}^{(r)}=0.
}
are used. From the relation
\eqn\conct{
\left[ \sum_{r<s}\theta^{ij}p_i^{(r)}p_j^{(s)},\
{X^i}^{(t)}(\sigma)-{X^i}^{(t+1)}(\pi-\sigma) \right]=
i\theta^{ij} p_j^{(t+2)}, 
}
the above equation \newconct\ leads us to find the three-string vertex 
with non-zero $B$-field
\foot{A similar expression for the three-string vertex
has been discussed in \FS.}

\eqn\vertex{
\V = \Vt \exp\left(-{i\o2} \sum_{r<s} \theta^{ij}
p_i^{(r)}p_j^{(s)} \right).
}

Thus, the three-string vertex in the background $B$ field 
can be obtained by multiplying the usual vertex $\Vt$ by the factor 
$e^{-{i\o2}\sum\theta^{ij}p_i^{(r)}p_j^{(s)}}$, 
which is characteristic of a noncommutative space.
Since the BRS charge $Q_{\rm B}$ can be expressed by the variables 
$\tilde{X}^i(\sigma)$ and $P_j(\sigma)$ and commute with the zero mode 
$p_j$ of the momenta, we can see that the three-string vertex satisfies 
the BRS invariance
\eqn\BRS{
\V \sum_{r=1}^3Q_{\rm B}^{(r)}=0.
}

Finally, we find that our string field theory has the following action:
\eqn\action{\eqalign{
S[\A]&=\int \left(\hf\,\A\,\star\,Q_{\rm B}\,\A
+{1\o3}\,\A\,\star\,\A\,\star\,\A\right) 
\cr
&=\hf{}_{21}\left\langle R\right|\left|\A\right\rangle_1 Q_{\rm B}^{(2)}
\left|\A\right\rangle_2
+{1\o3}{}_{321}\V
\left|\A\right\rangle_1\left|\A\right\rangle_2\left|\A\right\rangle_3.
\cr}
}
This $\star$ product is different from the ordinary 
product by the factor which represents the noncommutativity of space-time, 
as we have mentioned above. But, except for this factor, the action \action\ 
is the same as that of the theory without $B$-field.
Note that, in the kinetic term, the ordinary product can be replaced 
by the $\star$ product due to the momentum conservation. 
This action can be verified to be invariant under the gauge transformation
\eqn\gaugetrf{
\delta \A = Q_{\rm B}\Lambda+\A\star\Lambda-\Lambda\star\A.
}

In the perturbative expansion of this string field theory, 
if we expand the string field $\Psi$ by its component fields, 
for example, a tachyon field and a vector field, 
the product of these component fields in the resulting effective 
action turns out to be the product of functions on a noncommutative space. 
Therefore, the low-energy effective theory becomes noncommutative Yang-Mills 
theory. Also, in this theory, we have the open string metric $G_{ij}$, 
but not the closed one $g_{ij}$. 
This is in agreement with the result in \SW.


\newsec{Background Independence of String Field Theory}

As a background independent formulation of string theory, 
pregeometrical string field theories have been proposed 
in \refs{\HLRS,\HIKKOpre}, where they dropped the kinetic terms 
from the actions of the ordinary string field theories on a flat Minkowski 
space and kept only a cubic term. 

If we drop the kinetic term of our string field theory, 
we may expect the resulting theory to be a pregeometrical theory on 
the same footing as the theory proposed in \HLRS.
In this section, we will show that this is the case.
Although it seems to depend on the theta parameter $\theta^{ij}$, 
we will find that its background dependence can be absorbed into
a redefinition of a string field and that the resulting theory turns out 
to be the theory in \HLRS. In addition, we will explicitly show 
in the oscillator representation that the three-string vertex is also 
independent of $G_{ij}$. This is an application of the method given 
by Kugo and Zwiebach \KugoZwie\ to Witten's open string field theory.

In \KugoZwie, background independence has been discussed
in $\alpha=p^+$ closed HIKKO theory compactified on a torus. 
Kugo and Zwiebach proposed 
that $X^i(\sigma)$ and $P_i(\sigma)$ are independent of background fields. 
We can therefore read the dependence of the oscillators on the background 
fields, which allows us to explicitly verify in terms of the oscillators 
that the three-string vertex is background independent.

However, in our open string field theory, the coordinates $X^i(\sigma)$
are no longer universal objects, because the commutation relation
of string coordinates itself depends on $\theta$ as in \CCR. What objects
should we regard as universal ones? 
From Eq.~\Xmode\ and \qp, $X^i(\sigma)$ can be rewritten
as
$$
X^i(\sigma)=\tilde{X}^i(\sigma)+\int_{\pi\o2}^\sigma d\sigma'
\theta^{ij}P_j(\sigma')+\hf\theta^{ij}\left(
{p_L}_j-{p_R}_j\right).
$$
Thus, $X^i(\sigma)$ and $P_i(\sigma)$ can be expressed by
$\tilde{X}^i(\sigma)$
and $P_i(\sigma)$. Furthermore, their commutation relations 
\eqn\CCRR{
\left[\tilde{X}^i(\sigma),\ P_j(\sigma')\right]=i\,
\delta^i_j\,\delta(\sigma-\sigma'), 
\qquad
\left[ P_i(\sigma),\ P_j(\sigma') \right]=0, 
\qquad
\left[\tilde{X}^i(\sigma),\ \tilde{X}^j(\sigma')\right]=0.
}
have no apparent dependence on background fields. Thus, we 
propose that $\tilde{X}^i(\sigma)$ and $P_i(\sigma)$ are background independent 
objects. Namely, under  an infinitesimal variation of $G^{ij}$ and
$\theta^{ij}$, $\delta \tilde{X}^i(\sigma)=0$ and $\delta P_i(\sigma)=0$. 
Therefore, under the variation, we can obtain the change of 
the oscillators 
\eqn\da{
\delta\alpha_n^i=-\hf G^{ij}\delta G_{jk}\left(\alpha_n^k+\alpha_{-n}^k\right).
}
The oscillators $\alpha_n^i$ can be seen to only depend upon the 
open string metric $G^{ij}$.
This means that the theta parameter $\theta^{ij}$ in our theory is only 
included in the above-mentioned factor of the three-string vertex.

Now, let us consider a purely cubic action with our three-string vertex
\eqn\cubicaction{
S=\int \A\star\A\star\A
=  {}_{321}\V \ket{\A}_1\ket{\A}_2\ket{\A}_3.
}
If we expand the string field around a classical solution
as $\A=Q_{\rm L} I + \tilde{\A}$,
we can recover the action Eq.~\action, as discussed in \HLRS.
Here, $Q_{\rm L}$ is the BRS charge density integrated over the left
half of a string, and, in terms of the oscillators,
$I$ can be given \refs{\Samuel,\GJ} by
\eqn\identity{
\ket{I}=\exp\left(-\sum_{n\geq 1} {(-1)^n\o2n}
G_{ij}\alpha_n^i \alpha_n^j\right)\ket{0}(2\pi)^{26}\delta^{26}(p).
}

In the following two subsections, we will in turn discuss the dependence 
of our theory \cubicaction\ on the theta parameter $\theta^{ij}$ and 
on the open string metric $G_{ij}$. 

\subsec{Similarity Transformation of String Fields and the Theta Parameter}

The three-string vertex in our theory differs from that in \HLRS\ by the 
noncommutative factor $\exp[-(i/2)\sum_{r<s}\theta^{ij}p^{(r)}_ip^{(s)}_j]$.
Nothing but this factor depends on the theta parameter, as we have seen 
previously. Therefore, our theory at first seems dependent 
on the background field $\theta^{ij}$. If we are dealing with a particle field 
theory, say $\phi^3$ theory, on a noncommutative space, this is true. 
However, if, in our string theory, we can express the noncommutative factor 
by a product of operators from each of the three strings, we can eliminate 
the factor by a redefinition of string fields. Interestingly, this is indeed 
the case, as we will show below. Therefore, our pregeometrical theory 
is independent of the theta parameter.

To this end, let us consider the operator 
$\sum_{r<s} -(i / 2)\theta^{ij}p_i^{(r)}p_j^{(s)}$ on the ordinary 
three-string vertex $\Vt$. This operator can be rewritten as 
$\sum_{r=1}^3 (i / 2)\theta^{ij}{p_L}_i^{(r)}{p_R}_j^{(r)}$ 
by using the momentum conservation \PLPR\ on the vertex $\Vt$.
Therefore, our three-string vertex $\V$ can be rewritten as
\eqn\MVV{
\Vt \prod_{r=1}^3 e^{M^{(r)}}=\V,
}
where $M^{(r)}=(i / 2)\theta^{ij}{p_L}_i^{(r)}{p_R}_j^{(r)}$.

Since the noncommutative factor can be given by the product of the operators
$e^{M^{(r)}}$ on the three-string vertex, we can eliminate it from the vertex 
by a redefinition of string fields $\Psi\,\rightarrow\,e^{-M}\Psi$, and we find 
that our theory turns into the ordinary theory proposed by \HLRS, as we have 
mentioned before.

Before examining the dependence of our theory on the open string 
metric, we would like to make some comments about this similarity 
transformation. When we apply this field redefinition to the theory we have 
discussed in the last section, we can eliminate the noncommutative factor 
from the three-string vertex. But this redefinition also affects the kinetic 
term, and then the BRS charge is transformed into $e^{M}Q_{{\rm B}}e^{-M}$.
This transformed BRS charge can be found to have a divergent term in it. 
Very recently, using an interesting technique, Sugino has argued that 
the transformed operator indeed remains the original BRS operator 
$Q_{{\rm B}}$ in the kinetic term \Sugino. 
This seems to imply that the background $B$ field is physically 
meaningless. We would like to discuss this puzzle in some detail in section 4.

\subsec{Independence of Three-String Vertex from Background Metric}

In the string field theory, the reflector and the three-string vertex 
are defined by the overlap conditions up to an overall normalization. 
Since the overlap conditions do not include any background fields, 
we can expect that those vertices are independent of background fields. 
But we need at least a background metric to concretely construct those vertices 
in terms of the oscillators. Therefore, it is interesting to examine background 
independence of the vertices. 
In this subsection, we will consider the independence of the ordinary 
three-string vertex $\Vt$ from a background metric by using the method 
given by Kugo and Zwiebach, who applied it to the $\alpha'=p^+$ HIKKO closed 
string theory. We could also study the background independence by 
using a general method given by Sen \Sen.

Before considering the three-string vertex, we will demonstrate 
the independence of the reflector from the open string metric, 
as an illustration of the method of \KugoZwie. 
In this subsection, we will focus only on the matter sector.

The Fock vacuum of string fields is defined by 
${}_{G}\bvac\a_{-n}=0$ for $n\geq1$, where the oscillators $\a_{n}$ 
depend on the open string metric $G_{ij}$. Thus, the vacuum ${}_{G}\bvac$ 
also depends on the metric $G_{ij}$. As we have seen in \da, 
the oscillators change under an infinitesimal variation of $G_{ij}$ by
$\delta\alpha_n^i=-\hf G^{ij}\delta G_{jk}\left(\alpha_n^k+
\alpha_{-n}^k\right)$. It is useful to introduce an operator 
\eqn\Bogo{
{\cal B}=-\sum_{n\geq 1}{1\o4n}\delta G_{ij}
\left(\alpha_n^i \alpha_n^j-\alpha_{-n}^i \alpha_{-n}^j \right),
}
which satisfies $\left[ {\cal B},\,\alpha_n^i\right]=-\hf G^{ij}
\delta G_{jk} \alpha_{-n}^k$. According to the above definition of 
the Fock vacuum, it is changed under the variation $\delta G_{ij}$ into 
\eqn\vac{
{}_{G+\delta G}\left\langle0\right|
={}_G\left\langle0\right|-{}_G\left\langle0\right|{\cal B}.
}

The part of the reflector relevant to this paper is 
$\left\langle R^x\right| \sim 
{}_{21}\left\langle 0\right| e^{E_{12}}$, where $E_{12}$ is given 
\GJ\ by
$E_{12}=-\sum_{n\geq 1}{(-)^n \o n}G_{ij}{\alpha_n^i}^{(1)}{\alpha_n^j}^{(2)}$ 
and we will omit the delta function of the zero mode $p_j$, which, as we have 
mentioned before, is background independent. By making use of \vac\ and 
a formula $\delta\left( e^{E_{12}} \right)=\left[{\cal B}^{(1)}
+{\cal B}^{(2)},\, e^{E_{12}} \right]$ under the variation, we obtain 
\eqn\varref{
\delta \left\langle R^x\right| = -\left\langle R^x\right|
\left({\cal B}^{(1)}+{\cal B}^{(2)} \right).
}
The right-hand side of \varref\ is vanishing, because, on the reflector, 
the oscillators satisfy
$$
\left\langle R^x\right|\left({\alpha_n^i}^{(1)}+(-)^n {\alpha_n^i}^{(2)}\right)=0.
$$
Thus, the reflector is independent of the background $G_{ij}$.

Similarly, under the variation of the metric, 
we find the variation of the three-string vertex $\Vt$ to be
\eqn\variationA{
\delta \left\langle\tilde{V}_3^x\right| =
-\,{}_{321}\bra{0} e^{E_{123}} \sum_{r=1}^3 {\cal B}^{(r)}
+\,{}_{321}\bra{0}e^{E_{123}} \delta_0 E_{123},
}
where $\delta_0 E_{123}$ corresponds to the change in the zero-mode parts and 
is given by
$$
\delta_0 E_{123} = -\hf\sum_{rs}\bar{N}^{rs}_{00}
\delta G_{ij}{\alpha_0^i}^{(r)} {\alpha_0^j}^{(s)}
-\hf\sum_{rs}\sum_{m \geq 1}\bar{N}^{rs}_{0m}
\delta G_{ij}{\alpha_0^i}^{(r)} {\alpha_m^j}^{(s)}.
$$
The first term of \variationA\ can be evaluated to be 
${}_{321}\bra{0}e^{E_{123}}$ multiplied by
\eqn\variationB{\eqalign{
&-{1\o4}\sum_{n\geq 1}\sum_{r=1}^3{1\o n}
\delta G_{ij} {\alpha_n^i}^{(r)} {\alpha_n^j}^{(r)}
+{1\o4}\sum_{\sitarel{m,l\geq 1}{r,s}}
\left(\sum_{n\geq 1}\sum_{t=1}^3 \bar{N}_{mn}^{rt}n\bar{N}_{nl}^{ts}
\right)\delta G_{ij} {\alpha_m^i}^{(r)} {\alpha_l^j}^{(s)} 
\cr
&
-\hf\sum_{\sitarel{l\geq 1}{r,s}}
\left(\sum_{n\geq 1}\sum_{t=1}^3 \bar{N}_{0n}^{rt}n\bar{N}_{nl}^{ts}
\right)\delta G_{ij} {\alpha_0^i}^{(r)} {\alpha_l^j}^{(s)} 
-{1\o4}\sum_{r,s}
\left(\sum_{n\geq 1}\sum_{t=1}^3 \bar{N}_{0n}^{rt}n\bar{N}_{0l}^{ts}
\right)\delta G_{ij} {\alpha_0^i}^{(r)} {\alpha_0^j}^{(s)} 
\cr
&
-{1\o4}\sum_{n\geq 1}\sum_{r=1}^N n\bar{N}_{nn}^{rr}
\delta G_{ij} G^{ij}.
\cr}
}
Note that, although our argument is parallel to that in \KugoZwie, 
the last term of \variationB\ is a new term, of which we do not have 
the counterpart in the closed string case.

To prove the background independence, we can use the identities of 
the Neumann coefficients \refs{\GJ,\Yoneya},
\eqn\NeuIdA{\eqalign{
\sum_{n\geq 1}\sum_{t=1}^3 \bar{N}_{mn}^{rt}n\bar{N}_{nl}^{ts}
&={1 \o m}\delta_{m,l}\delta^{rs},
\qquad
\sum_{n\geq 1}\sum_{t=1}^3 \bar{N}_{mn}^{rt}n\bar{N}_{n0}^{ts}
=-\bar{N}^{rs}_{m0},
\cr
&\sum_{n\geq 1}\sum_{t=1}^3 \bar{N}_{0n}^{rt}n\bar{N}_{n0}^{ts}
=-2\bar{N}_{00}^{rs}.
\cr}
}
Note that the second and third equalities need the momentum conservation 
for the zero-modes, which is guaranteed by the vertex. 
These identities are proven in appendix B. 
Therefore, we can see 
that the second term of \variationA\ cancels the first three terms of 
\variationB. For the last term of \variationB, we need another identity 
\eqn\NeuIdB{
\sum_{n\geq 1}\sum_{r=1}^N n\bar{N}_{nn}^{rr}=0,
}
which is also proved in appendix B. Thus, we can see that 
the three-string vertex $\Vt$ is background independent.


\newsec{Discussion}

In this paper, we derived Witten's open string field theory in a background 
$B$ field by using the standard overlap conditions in the operator formalism.
The resulting three-string vertex naturally contains an additional factor 
which gives the Moyal product to the zero modes, compared to the ordinary 
vertex with no $B$ field. Thus, the zero modes $\tilde{x}^i$ can be found to be
noncommutative. Besides this noncommutative factor, the three-string vertex can 
be written by using the open string metric $G_{ij}$. Therefore, the low-energy 
effective theory of the gauge field should be described by noncommutative 
Yang-Mills theory. This result is in agreement with the result in the 
first-quantization formulation in \refs{\SW,\Schom}.

Following the idea of the pregeometrical formulation \refs{\HLRS,\HIKKOpre}, 
we dropped the kinetic term from our string field theory and explicitly 
demonstrated background independence of the resulting theory by using 
the method of \KugoZwie. 

In order to prove that the three-string vertex is independent of the theta 
parameter $\theta^{ij}$, we have shown that the noncommutative factor can 
be eliminated by field redefinition. If we also apply the redefinition to 
the theory with the kinetic term, we can easily see that, besides elimination 
of the noncommutative factor of the three-string vertex, it also affects 
the kinetic term and transforms the BRS charge $Q_{{\rm B}}$ into
$e^{M}Q_{{\rm B}}e^{-M}$. Here the operator $M$ is 
$(i / 2)\theta^{ij}{p_L}_i{p_R}_j$. This transformed BRS charge can be found 
to have a divergent term in it. This divergent term seems to come from 
the mid-point $\sigma=\pi/2$ of strings. 
Very recently, using an interesting technique, Sugino has argued \Sugino\ 
that the transformed operator indeed remains the original BRS operator 
$Q_{{\rm B}}$ in the kinetic term 
and that the kinetic term is kept intact under the field redefinition. 
This seems to imply that the background $B$ field is physically 
meaningless. 

In the paper \SW, Seiberg and Witten have shown that the noncommutative 
Dirac-Born-Infeld (DBI) action is equivalent in the slowly varying field 
approximation to the commutative DBI action with the background 
$B$ field. In addition, they emphasized that the $B$ field parallel to  
$D$-branes cannot be gauged away, due to the gauge invariance 
$B_{ij}\rightarrow B_{ij}+\d_i\Lambda_j-\d_j\Lambda_i$, 
$A_i \rightarrow A_i+\Lambda_i$. Here $A_i$ is the gauge field on the D-branes. 
Therefore, to be consistent with the result 
in \SW, we may expect that, even after the field redefinition, the dependence 
of the $B$ field remains in the kinetic term of the string field theory 
and that the $B$ field appears in the low-energy effective action only through 
the gauge-invariant combination ${\cal F}_{ij}=B_{ij}+F_{ij}$.
However, it apparently seems to conflict with Sugino's recent result \Sugino.
Thus, we think that we have an interesting puzzle to solve. 

To our knowledge, there is no literature which shows that Witten's open 
string field theory has the above-mentioned gauge invariance. 
In addition, it seems difficult to prove it, because the theory has no 
explicit field from the closed string sector in its action.
For this purpose, it may be more suitable to study the gauge invariance and 
the dependence of $B$ field in an open-closed string field theory 
with the mid-point interaction in \Wopenclosed.

Apart from the gauge invariance, we would like to discuss the operator 
$M$ which was used for the field redefinition. 
If we do not use Sugino's technique to show that the kinetic term 
remains intact by the field redefinition, we have to deal with the divergent 
term in $e^{M}Q_{{\rm B}}e^{-M}$. This divergent term seems to come from the 
mid-point of strings, as we mentioned above. Since the operator $M$ consists of 
the half-integrated momenta $p_L$, $p_R$, we are led to wonder if 
the singularity may be related to the mid-point interaction. 
In addition, the operator $M$ seems to be suitable only to the mid-point 
interaction, because we cannot apply it to the light-cone type interaction.
Since the kinetic term of string field theories does not depend on types of 
interactions of strings, it is desirable to have a field redefinition which 
is independent of types of string interactions. 
Therefore, it may be useful to introduce another candidate 
\eqn\M{
\tilde{M}=-{i\o4}\int_0^\pi d\sigma\int_0^\pi d\sigma'
\epsilon(\sigma-\sigma')\theta^{ij}P_i(\sigma)P_j(\sigma'),
}
where $\epsilon(\sigma)$ is the step function which
is $1$ for $\sigma>0$ and $-1$ for $\sigma<0$.
Indeed, rewriting this operator as
\eqn\Mrw{
{i \o 2}\theta^{ij}{p_L}_i{p_R}_j-{i\o 4}
\int_{\pi \o 2}^\pi d\sigma \int_{\pi \o 2}^\pi d\sigma'
\epsilon(\sigma-\sigma')\theta^{ij}
\{P_i(\sigma)P_j(\sigma')-P_i(\pi-\sigma)P_j(\pi-\sigma')\}
}
and putting the sum $\sum^{3}_{r=1}\tilde{M}^{(r)}$ 
on the usual three-string vertex $\Vt$, we can see that 
the second terms in the operators $\tilde{M}^{(r)}$ cancel 
each other, due to the overlap condition \VtX. 
Furthermore, as we will mention just below, 
this operator $\tilde{M}$ can be used to give the field redefinition 
to remove the noncommutative factor from the light-cone type interactions.

We have so far been discussing the problem with the mid-point interaction 
concerning the relation between our string field theory and the ordinary one. 
However, it is plausible that, if there is such a relation, we can 
find the same relation in other string field theories. 
In particular, since we cannot apply the operator $M$ to string field theories 
with the light-cone type interaction like that given by \AKT, we can expect 
that the dependence on the $B$ field cannot completely be removed. 
Therefore, in order to get some clue to this problem, 
it may be helpful to study the string field theory with the light-cone type 
interaction \AKT\ by using the operator $\tilde{M}$ . 
Since this theory explicitly has closed string fields in its Lagrangian 
as well as open strings, it may also help 
to find some relation between the condensation of the antisymmetric tensor
$B_{ij}$ from the closed string field and the above redefinition of the open 
string field. 

Now, let us just sketch in the main points of our results about the 
string field theory with the light-cone type interaction \AKT\ in 
the background $B$ field. These results will be explained in more detail 
in another paper \KT. As we can verify by the method we have explained 
in section 2, the light-cone type vertices are also modified to include 
the noncommutative factor, and there is no other modification due to 
the background $B$-field. This noncommutative factor can be expressed by a 
product of the operators $\tilde{M}$ from each string. Since the operator 
$\tilde{M}$ thus plays an important role, it is useful to make some comments
on it. 

The commutation relation between the operator $\tilde{M}$ and 
the string coordinates can be verified to be 
\eqn\Mx{\eqalign{
[\tilde{M},\,{X}^i(\sigma)]
=& -{1 \o \pi^2 l_s}\sum_{n\neq 0}{1-(-)^n \o n^2} (\theta G)^i_j
\alpha_n^j
+{2 \o \pi^2}\theta^{ij}\,p_j\sum_{m\geq 1}{1-(-)^m \o m^2}\cos(m\sigma)
\cr
&+{2 \o \pi^2 l_s}\sum_{m\geq 1}\left(
\sum_{n\neq \pm m}{1-(-)^{m+n} \o m^2-n^2}(\theta G)^i_j \alpha_n^j
\right) \cos(m\sigma).
}}
If we naively exchange the order of the summations in the right-hand side 
of \Mx, we obtain $[\tilde{M},\,{X}^i(\sigma)]=-(\theta G)^i_j\, Q^j(\sigma)$. 
Therefore, we would find that $e^{\tilde{M}}\,X^i(\sigma)\,e^{-\tilde{M}}=
\tilde{X}^i(\sigma)$.
But this consequence must be false because it is inconsistent with the 
commutation relation $[{X}^i(\sigma),\ {X}^j(\sigma')]\sim\theta^{ij}$, 
which cannot be changed to be $[\tilde{X}^i(\sigma),\ \tilde{X}^j(\sigma')]=0$ 
by the similarity transformation $e^{\tilde{M}}\,X^i(\sigma)\,e^{-\tilde{M}}$. 
A closer examination shows that the relation 
$[\tilde{M},\,{X}^i(\sigma)]=-(\theta G)^i_j\, Q^j(\sigma)$ holds 
only for $0<\sigma<\pi$. Therefore, the transformed BRS operator 
$e^{\tilde{M}}\,Q_{{\rm B}}\,e^{-\tilde{M}}$ naively becomes the ordinary 
BRS operator with the closed string metric $g_{ij}$, but, due to subtleties 
from the ends of strings, it has additional terms for which, at least at 
present, we do not have any interpretations. 
In spite of the discrepancy between 
$e^{\tilde{M}}\,X^i(\sigma)\,e^{-\tilde{M}}$ 
and $\tilde{X}^i(\sigma)$, since the operator $\tilde{M}$ allows us to 
relate our three-string vertex to the ordinary vertex, regardless of types of 
string interactions, we are tempted to speculate that the operator $\tilde{M}$ 
would give us some clue about a relation between the noncommutative string 
field theory and the ordinary one, like the relation found by Seiberg and 
Witten \SW\ between noncommutative DBI theory and commutative one.

Finally, we would like to touch on superstring field theory. 
We can easily extend our theory to Witten's superstring field theory 
\WittenSSFT\ by constructing other necessary vertices in a similar way 
to the bosonic case.
In the worldsheet picture of superstring theory,
we add to the bosonic sector
\eqn\superaction{
S_{\psi} = -{1\o4\pi\alpha'}
\int d^2 z \left( g_{ij}\psi^i \bar{\partial} \psi^j
+ g_{ij}\tilde{\psi}^i \partial \tilde{\psi}^j \right),
}
and the boundary conditions are given by
\eqn\superbc{
(g+2\pi \alpha' B)_{ij} \psi(z) =
(g-2\pi \alpha' B)_{ij} \tilde{\psi}(\bar{z}),\ \ \ {\rm at}\ z=\bar{z}.
}
It is convenient to introduce new fields $\varphi^i(z)$, $\tilde{\varphi}^i(z)$ 
defined by 
\eqn\interfermion{\eqalign{
&\varphi^i(z) =G^{ij}(g+2\pi\alpha'B)_{jk}\psi^k(z)
\cr
&\tilde{\varphi}^i(\bar{z})=
G^{ij}(g-2\pi\alpha'B)_{jk}\tilde{\psi}^k(\bar{z})
\cr}
}
These fields play a similar role to the string coordinates 
$\tilde{X}^i$ in the bosonic case. Since the fields $\varphi^i(z)$ 
and $\tilde{\varphi}^i(z)$ satisfy 
the same boundary condition as that with no $B$ field, we have the ordinary 
mode expansion of these fields $\varphi^i(z)$, $\tilde{\varphi}^i(z)$. 
Therefore, solving the overlap conditions for these fields, we obtain 
the ordinary three-string vertex as well as the ordinary reflector. 
The oscillator expression of the vertices can be found in \refs{\Suehiro,\GJ}. 
From \interfermion, we can immediately verify that these vertices also 
satisfy the overlap conditions for $\psi(z)$, $\tilde{\psi}(\bar{z})$. 
Moreover, the picture changing operators can be expressed by using only 
$\varphi^i(z)$, $\tilde{\varphi}^i(z)$ and $\tilde{X}^i$, and can be found 
to have the ordinary expression.
Thus, we can extend our string field theory to superstring 
cases, though we are still faced with the mid-point
singularity problem as in \Wendt.\foot{
Recently, a new approach has been proposed to solve
this difficulty \Berkovits.}


\medskip
\centerline{{\bf Acknowledgements}}
The authors would like to thank Tsuguhiko Asakawa, Katsumi~Itoh, Taichiro~Kugo, 
Masahiro~Maeno, Kazumi Okuyama, Kazuhiko Suehiro, and Seiji Terashima 
for valuable discussion. 
T.$\,$K. is grateful to 
the organizers and the participants of Summer Institute '99 in Yamanashi, Japan
for the hospitality and stimulating atmosphere created there, which helped 
to initiate this work. 
T.$\,$K. was supported in part by a Grant-in-Aid (\#11740143)
and in part by a Grant-in-Aid for Scientific Research 
in a Priority Area: ``Supersymmetry and Unified Theory of Elementary 
Particles''(\#707), from the Ministry of Education, Science, Sports and 
Culture.
T.$\,$T. is supported in part by Research Fellowships of the Japan Society for
the Promotion of Science for Young Scientists.


\appendix{A}{Operator Formalism of Strings in a Background $B$-Field}

We consider the operator formalism of first-quantized string theory 
in a constant background $B$ field by following the papers 
\refs{\ChuHo,\SJabbari}.
Although Chu and Ho have discussed different methods of quantization 
in their first and second papers of \ChuHo,
our strategy is slightly different from both of them; namely, 
we simplify their methods by combining them.

The worldsheet action is given by
\eqn\WSaction{
S=-{1\over4\pi\alpha'}\int d\sigma\,d\tau
\left(g_{ij}\eta^{ab}\partial_a X^i \partial_b X^j
-2\pi\alpha'B_{ij}\epsilon^{ab}
\partial_a X^i \partial_b X^j \right),
}
where $g_{ij}$ is a constant background metric and $B_{ij}$ is a constant 
background antisymmetric tensor. We will also denote $(2\pi\a')B_{ij}$ 
by $b_{ij}$. The signature of the worldsheet metric 
$\eta^{ab}$ is $(-,+)$ and the invariant antisymmetric tensor $\epsilon^{ab}$ 
is defined by $\epsilon^{01}=1$. The equation of motion of the string 
coordinates is $\d_{a}\eta^{ab}\d_{b}X^{i}=0$. The boundary condition turns out 
to be
\eqn\bc{
g_{ij}{X'}^j+ 2\pi\alpha' B_{ij}\dot{X}^j =0
}
at $\sigma=0,\,\pi$. The conjugate momenta are given by 
$P_i = (1/2\pi\alpha')g_{ij}\dot{X}^j + B_{ij}{X'}^j$, where 
$\dot{}\ $ and $'$ denote the differentiation with respect to 
$\tau$ and $\sigma$, respectively. As in the usual way, we can obtain the 
Hamiltonian density 
\eqn\Hamilton{
{\cal H}={1\o4\pi\a'}\left[
(2\pi\a')^2g^{ij}P_iP_j+(4\pi\a')b_{ik}g^{kj}{x'}^iP_j+G_{ij}{X'}^i{X'}^j
\right].
}
Here, the open string metric $G_{ij}$ is given by 
$G_{ij}=g_{ij}-(bg^{-1}b)_{ij}$.
The Poisson brackets of the string coordinates and the momenta are 
all vanishing except $\left\{X^i(\sigma),\,P_{j}(\sigma')\right\}_P=
\delta^i_j\delta(\sigma-\sigma')$.

The idea in the second paper of \ChuHo\  and in \SJabbari\  was to deal with 
the boundary condition \bc\ as a constraint 
$\phi_i=G_{ij}{X'}^j+(2\pi\a')b_{ik}g^{kj}P_{j}$ in the operator formalism 
and to quantize the system by following the Dirac quantization method.
By using the consistency condition 
$\dot{\phi_i}=\left\{\phi_i, H\right\}_P\approx0$ 
with the Hamiltonian $H=\int d\sigma{\cal H}$, all the second class constraints 
are found \refs{\ChuHo,\SJabbari} to be 
\eqn\constraint{
{\d^{2n}\phi_i\o\d\sigma^{2n}}(\sigma)=0,
\qquad
{\d^{2n+1}P_i\o\d\sigma^{2n+1}}(\sigma)=0,
}
with $n\geq0$ at $\sigma=0, \pi$, and there is no first class constraint. 
Here, we first solve these constraints \constraint\ and find that
\eqn\constsol{
\phi_i(\sigma)=-\sum^{\infty}_{n=1}nG_{ij}x^j_{n}\sin(n\sigma),
\qquad
P_i(\sigma)=\sum^{\infty}_{n=0}p_{ni}\cos(n\sigma).
}
Therefore, using $\theta^{ij}=-(2\pi\a')(G^{-1}bg^{-1})^{ij}$, we obtain 
\eqn\mode{
X^i(\sigma)=\sum^{\infty}_{n=0}x^{i}_{n}\cos(n\sigma)
+\theta^{ij}\left[p_{0j}\sigma
+\sum^{\infty}_{n=1}{1\o n}p_{nj}\sin(n\sigma)\right].
}

Chu and Ho in their first paper of \ChuHo\ expressed the string coordinates 
as a mode expansion in terms of the solutions of the equation of motion 
and used the invariant symplectic form
\eqn\symplectic{
\omega=\int d\sigma \left[-dX^i(\sigma)\wedge dP_i(\sigma)
+dP_i(\sigma)\wedge dX^i(\sigma)\right]
}
to find the commutation relations of the modes. 
Now we apply their idea to our system 
by using the above mode expansion of string coordinates and the momenta, 
instead of the mode expansion in terms of the solutions of the equation of 
motion, to find the commutation relations of $x^i_n$ and $p_{nj}$.
After this procedure, we will find the equation of motion of these variables
$x^i_n$ and $p_{nj}$. Substituting the expansions \constsol\ and \mode\ into 
the symplectic form \symplectic, we find the Poisson brackets of the modes to be
\eqn\Poisson{
\left\{x^i_0, p_{0j}\right\}_P={1\o\pi}\delta^i_j,
\qquad
\left\{x^i_n, p_{mj}\right\}_P={2\o\pi}\delta^i_j\delta_{n,m},
\qquad
\left\{x^i_0, x^j_0\right\}_P=\theta^{ij},
}
and the others are vanishing. Since these variables $x^i_n$ and $p_{nj}$ are 
the solutions of the constraints \constraint, they are all physical variables. 
Thus, we can obtain the commutation relations of them by the usual prescription 
$[A, B]=i\{A, B\}_P$ to quantize our system.

Since the time derivative of a physical variable ${\cal O}$ can be obtained 
by $\dot{{\cal O}}=\left\{{\cal O}, H\right\}_P$, we can see that
$\dot{\phi_i}(\sigma)=(2\pi\a'){P'}_i(\sigma)$, 
$\dot{P}_i(\sigma)=(1/2\pi\a'){\phi'}_i(\sigma)$.
Substituting the mode expansion \constsol\ into these equations to get 
the equations of motion of the variables $x^i_n$ and $p_{nj}$
and solving the resulting equations, we find that 
\eqn\amode{
x^i_n=i{l_s \o n}\left(\a^i_n e^{-in\tau}-\a^i_{-n} e^{in\tau}\right),
\qquad
p_{nj}={1\o\pi l_s}G_{jk}\left(\a^k_n e^{-in\tau}+\a^k_{-n} e^{in\tau}\right),
}
with $l_s=\sqrt{2\a'}$, for $n\not=0$. 
We also have $x^i_0=x^i+{l_s}^2G^{ij}p_{j}\tau$ and $p_{0j}=(1/\pi)p_{j}$.
Putting these new mode variables $\a^i_n$, $x^i$, $p_i$ into \Poisson, 
we obtain 
\eqn\Commutation{
\left[x^i, p_j \right]=i\delta^i_j,
\qquad
\left[x^i, x^j \right]=i\theta^{ij},
\qquad
\left[\a^i_m, \a^j_n \right]= mG^{ij}\delta_{m+n}.
}
These commutation relations are in agreement with the results in \ChuHo.
Note here that, if we define new `center of mass' coordinates $\tilde{x}^i$ 
by $\tilde{x}^i=x^i+(1/2)\theta^{ij}p_j$, 
the coordinates $\tilde{x}^i$ turn out to be commutative variables;
$\left[\tilde{x}^i, \tilde{x}^j \right]=0$.

Finally, with the mode variables $\a^i_n$, $\tilde{x}^i$, $p_i$, 
we can express the string coordinates $X^i(\sigma)$ and 
the conjugate momenta $P_i(\sigma)$
\eqn\modeexp{\eqalign{
&X^i(\sigma)=\tilde{X}^i(\sigma)
+{1\o\pi l_s}\theta^{ij}\left[l_sp_j\left(\sigma-{\pi\o2}\right)
+G_{jk}\sum_{n\not=0}{1\o n}\a^k_ne^{-in\tau}\sin(n\sigma)\right],
\cr
&P_j(\sigma)={1\o\pi l_s}G_{jk}\sum^{\infty}_{n=-\infty}
\a^k_ne^{-in\tau}\cos(n\sigma),
\cr}
}
where $\tilde{X}^i(\sigma)=\tilde{x}^i+{l_s}^2G^{ij}p_j\tau
+l_s\sum_{n\not=0}(i/n)\a^i_ne^{-in\tau}\cos(n\sigma)$ and 
$p_j=(1/l_s)G_{jk}\a^k_0$.


\appendix{B}{Identities of the Neumann coefficients}

Consider the $N$-strings vertex with a midpoint interaction 
\refs{\GJ,\CST,\Samuel}.
The Neumann function on the strip is given by
\eqn\Neumann{\eqalign{
N(\rho_r, \rho'_s) &= -\delta_{rs}\left[2\sum_{n\geq 1}{1 \o n}
e^{-n|\tau_r-\tau'_s|}\cos(n\sigma_r)\cos(n\sigma'_s)-2\max(\tau_r,\tau'_s)
\right]
\cr
& +2\sum_{m,n\geq 0} \bar{N}_{mn}^{rs}e^{m\tau_r+n\tau'_s}
  \cos(m\sigma_r)\cos(n\sigma'_s).
\cr}
}
In the case of $\tau_r>\tau'_s$, we find that
$$\eqalign{
{\partial \o \partial\tau_r}
N(\rho_r, \rho'_s) &= 2\delta_{rs}\left[\sum_{n\geq 1}
e^{-n|\tau_r-\tau'_s|}\cos(n\sigma_r)\cos(n\sigma'_s)+1
\right]
\cr
&+2\sum_{{m\geq 1}}\sum_{{n\geq 0}}
m\bar{N}_{mn}^{rs}e^{m\tau_r+n\tau'_s}
  \cos(m\sigma_r)\cos(n\sigma'_s).
\cr}
$$
The Neumann function and its $\rho$ derivative are continuous
at the interaction time $\tau_r=0$, provided that we use
momentum conservation for the zero mode parts. In other words,
in order to hold its continuity, it is necessary to multiply the factor
$\sum_s p^{(s)}$ to the zero mode terms $2\delta_{rs}\max(\tau_r,\tau'_s)$
and $\bar{N}_{n0}^{rs}$.

Using the continuity of the Neumann function, we find
the following identity,
$$\eqalign{
0&=\sum_{t=1}^N \int_0^\pi d\sigma''_t N(i\sigma''_t, \rho_r)\,
{\partial N\o\partial\tau''_t}(i\sigma''_t,\rho'_s)
\cr
&= -2\pi\delta_{rs}\sum_{n\geq 1}{1 \o n}e^{n\tau_r+n\tau'_s}
\cos(n\sigma_r)\cos(n\sigma'_s)
\cr
&+2\pi\sum_{n\geq1}\bar{N}^{rs}_{n0}e^{n\tau_r}\cos(n\sigma_r)
+2\pi\sum_{n\geq1}\bar{N}^{rs}_{0n}e^{n\tau'_s}\cos(n\sigma'_s)
+4\pi\bar{N}^{rs}_{00} 
\cr
&+2\pi \sum_{m,l\geq 1}
\left(\sum_{n\geq 1}\sum_{t=1}^N \bar{N}_{mn}^{rt}n\bar{N}_{nl}^{ts}\right)
e^{m\tau_r+l\tau'_s}\cos(m\sigma_r)\cos(l\sigma'_s)
\cr
&+2\pi\sum_{m\geq 1}
\left(\sum_{n\geq 1}\sum_{t=1}^N \bar{N}_{mn}^{rt}n\bar{N}_{n0}^{ts}\right)
e^{m\tau_r}\cos(m\sigma_r)
\cr
&+2\pi\sum_{l\geq 1}
\left(\sum_{n\geq 1}\sum_{t=1}^N \bar{N}_{0n}^{rt}n\bar{N}_{nl}^{ts}\right)
e^{l\tau'_s}\cos(l\sigma'_s).
\cr}
$$
From this identity, we can find \NeuIdA.

To obtain \NeuIdB, 
we take the limit $\rho'_s\rightarrow\rho_r$ in the Neumann
function \Neumann.
$$\eqalign{
N(\rho_r, \rho_r+\delta) &= \ln\delta
-\sum_{n\geq 1}{1 \o n}\cos(2n\sigma_r)+2\tau_r 
\cr
& +2\sum_{m,n\geq 0} \bar{N}_{mn}^{rs}e^{(m+n)\tau_r}
  \cos(m\sigma_r)\cos(n\sigma_r)+{\rm O}(\delta^2).
cr}
$$
From the continuity at the time of interaction,
we can obtain \NeuIdB\ as follows,
$$
0=\sum_{r=1}^N\int_0^\pi d\sigma_r {\partial \o \partial\tau_r}
N(i\sigma_r,i\sigma_r+\delta) 
= 2\pi \sum_{n\geq 1}\sum_{r=1}^N n \bar{N}_{nn}^{rr},
$$
where we can vanish $\sum_{r=1}^N
\partial/{\partial \tau_r}\,\tau_r$ due to the momentum
conservation.

These arguments and identities can also be established
for the light-cone type vertices with one interaction time.

\listrefs

\end